\documentclass[aps,showpacs,superscriptaddress,preprint,amsmath,amssymb]{revtex4}
\usepackage{graphicx}
\usepackage{dcolumn}
\usepackage{bm}

\begin{document}
\title{ Spallation caused by the diffusion and agglomeration of vacancies in ductile metals}
\author{Yuanjie Huang} \affiliation{The National Key Laboratory of Shock Wave and Detonation Physics, Institute of Fluid Physics, Chinese Academy of Engineering Physics, Mianyang 621900, China}
\date{\today}

\begin{abstract}
In this work, the spallation processes in the ductile metals are systematically discussed in theory. By employing the phase transition theory and non-equilibrium transport theory, the spallation processes of ductile metals under dynamic loading may be mainly determined by the diffusion and agglomeration of generated vacancies. And through the theoretical analysis on the dynamic spallation processes, it is found that (1)the spallation critical behaviors exist; (2)both the damage evolution rate and the void growth velocity in the spallation planes are monitored by the grain sizes, the tensile strain rate and the temperature of the metal after shocking, i.e., a larger grain size and tensile strain rate and higher temperature will cause a larger damage evolution rate and void growth velocity; (3)there exists the characteristic size for the voids and the characteristic stress on the void boundary, which are dependent of vacancy excitation energy and the average volume occupied by the metal atom; (4)both the stress and temperature close to the void are high and may cause the melting, and they decrease quickly as the distance away from the void increases; (5)the plastic zone surrounding the formed voids is obtained and found to be governed by the characteristic stress; (6)the void growth in the spallation processes may arise from the agglomeration of vacancies rather than the emitting of dislocations. Most of the theoretical results are novel and obtained first. And they are found to be in agreement with the experimental results and the simulated results.

\noindent\emph{keywords}: spallation;\quad ductile metals;\quad damage;\quad void
\end{abstract}

\pacs{61.72.-y}

\maketitle

\noindent\textbf{1. Introduction}

Due to the wide spectrum of issues ranging across defense and industrial applications, material behaviors in high pressure, high strain and high strain-rate physical processes, e.g. spallation processes, is interesting in science. For the ductile metals in the spallation processes typically possessing the above extreme physical conditions, the dynamic damage will be caused and is experimentally found to exist in the form of voids with different sizes in the spallation planes\cite{meilan,Belak,xiaoyang}. And the void nucleation, void growth and void coalescence are thought to play the main role for the ductile metals in the spallation processes. Because of the importance of void dynamic behaviors in the spallation, the void sizes, void growth velocity, coalescence between two nearby voids, damage evolution rate and so on, have been investigated persistently in experiments and theory several decades\cite{meilan,Belak,xiaoyang,NAG,Gurson,Johnson,Curran,critical behavior,Reina}. However, up to date, owing to the limitations caused by the lacking of effective in-situ characterization within short time in experiments and the challenge mainly brought by the cross-scale dynamic properties of the spallation phenomena in space and time in theory, the corresponding spallation phenomena have not been understood well and the related effective theoretical treatment is still an open problem.

In this paper, the spallation processes in the ductile metals are systematically discussed in theory. In terms of the phase transition theory and non-equilibrium transport theory, the spallation processes of ductile metals under dynamic loading may be mainly determined by the diffusion and agglomeration of generated vacancies as revealed by the simulations\cite{Reina}. And through the theoretical analysis on the dynamic spallation processes, it is found that (1) the spallation critical behaviors exist; (2)both the damage evolution rate and the void growth velocity in the spallation planes are monitored by the grain sizes of polycrystal, the tensile strain rate and the temperature of the metal after shocking, i.e., a larger grain size and tensile strain rate and higher temperature will cause a larger damage evolution rate and void growth velocity; (3)there exists the characteristic size for the voids and the characteristic stress on the void boundary, which dependent on the vacancy excitation energy and the average volume occupied by the metal atom; (4)both the stress and temperature close to the void are high and may cause the melting, and they decrease quickly as the distance away from the void increases ; (5)the plastic zone surrounding the formed voids is obtained and found to be governed by the characteristic stress;(6)against the widely accepted viewpoint that the void growth is induced by the emitting of dislocation\cite{Belak,dengxiaoliang, zhuwenjun}, it is proposed and argued that the void growth in the spallation processes may arise from the coalescence of vacancies rather than the emitting of dislocations. They are compared with the experimental results and the simulated results, and the agreement is found between them. In all, the theories depict the spallation and the corresponding dynamic behaviors, and the results build the bridge between the microscopic and macroscopic behaviors during spallation processes.



\noindent\textbf{2. Theory}

\begin{figure}[B]
\includegraphics[scale=1.5,angle=0]{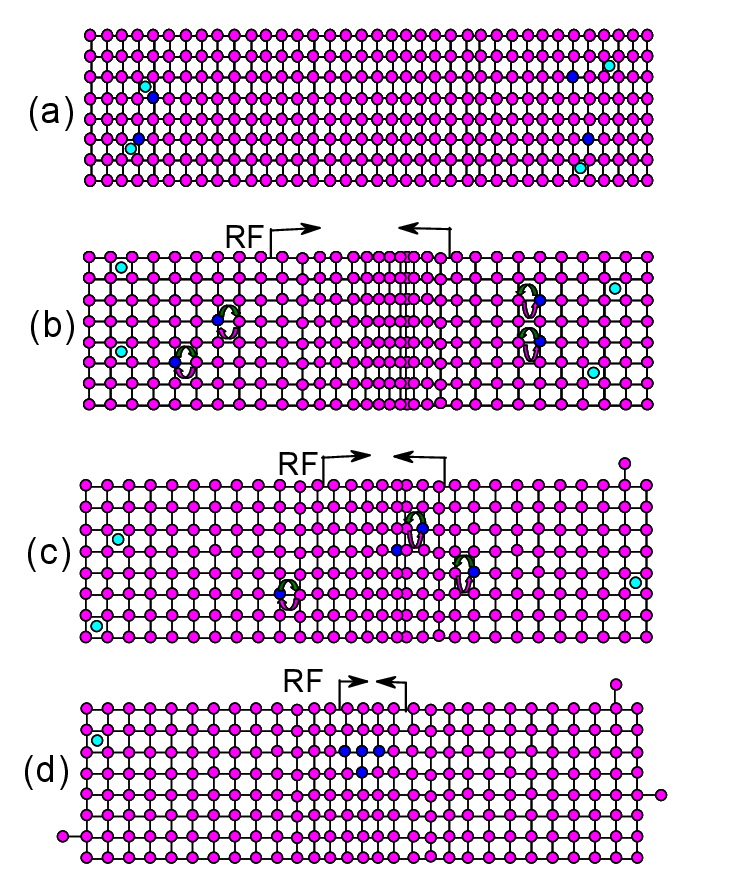}
\caption{Schematic diagrams of the sample in dynamic experiments: (a)the configurations of the atoms (pink circled), vacancies (blue circled) and interstitial atoms (light blue circled) in the metal after the shock loading; (b) the propagation of two series of rarefaction waves (RF) and the motion of vacancies and atoms; the black arrows indicate the propagation of RF; (c) the formation of small void caused by the agglomeration of vacancies in the spallation planes; (d) the further growth of void due to the coalescence of vacancies. }\label{RF}
\end{figure}

 Before the quantitative theory being presented, the roles of vacancies should be discussed qualitatively first. For the shock-loaded ductile monatomic metal, the temperature and stress are incremented greatly, and consequently a large number of vacancies may be yielded by the excitation of Frenkel defects, one vacancy and one concomitant interstitial atom (see Figure.\ref{RF}(a)). The vacancies may lie in the equilibrium state within the femtosecond time due to the related excitation energy $\sim eV$ and obey the distribution function $e^{-\varepsilon_{n}(\epsilon)/(2k_{B}T)}$, where $\varepsilon_{n}(\epsilon)$ is the strain $\epsilon$ dependence of vacancy excitation energy, $k_{B}$ is the Boltzmann constant, $T$ is the temperature under shock loading, and the factor 2 comes from the entropy increment. The yielded vacancies whose mobility may be much higher than that of interstitial atoms are randomly distributed until they enter the rarefaction zones. In the rarefaction zone, the vacancies move in the opposite directions with the atoms and go to the high stress zones such as the grain boundaries in the polycrystal, then reside there. The directional velocity of the vacancies is large and may achieve $\sim 10^{3}$ $m/s$. So the primarily randomly distributed vacancies will be pushed to a narrow region by the two series of rarefaction waves shown in Fig.\ref{RF}(b), where the concentration of vacancies is much higher than other regions. The vacancies may coalesce to form voids with different sizes (see Fig.\ref{RF}(c)), thereby leading to the spallation in these regions.


\noindent\textbf{A. vacancy-vacancy interaction}

To describe the spallation processes quantitatively for ductile metals, the phase transition theory for ferromagnetic transition in condensed matter physics are utilized. Wherein the interaction between any two vacancies is the key and should be constructed first.

\begin{figure}[B]
\includegraphics[scale=1,angle=-90]{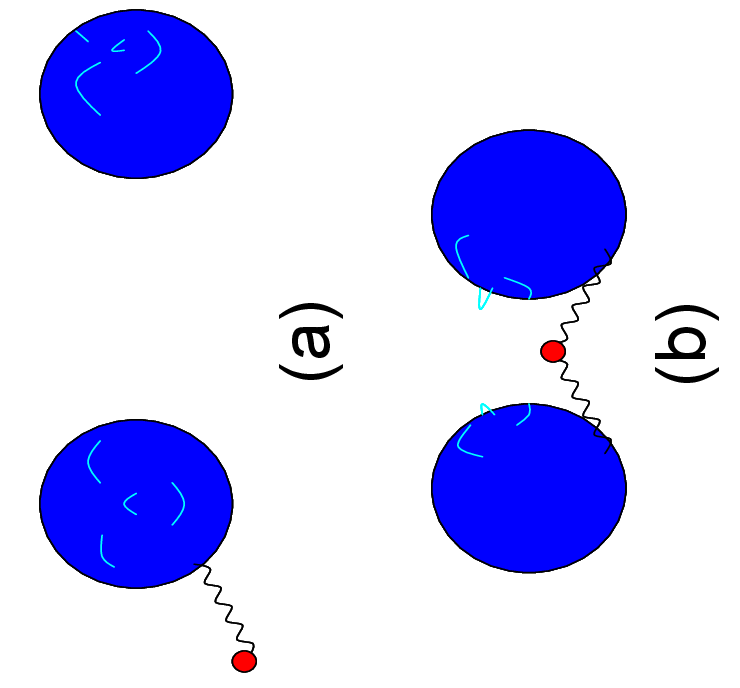}
\caption{Schematic diagrams of the interaction among voids (blue person): (a) non-interaction between the two voids; (b)the attraction induced by the electron (red circle) between the two voids.}\label{vv}
\end{figure}

For the ductile metals, one existing vacancy will alter the electronic potential field for the nearby itinerant electrons and result in the interaction $\hat{V}_{ev}$ between electrons and vacancies as shown in Fig.\ref{vv}(a). Let field variables $\Phi(\textbf{R})$,$\Psi(\textbf{r})$ denote the vacancy field and electron field, respectively, and the vacancy density gives  $\rho(\textbf{R})=\Phi^{\ast}(\textbf{R})\Phi(\textbf{R})=\sum_{i}\delta(\textbf{R}-\textbf{R}_{i})$.
The electron-vacancy (e-v) interaction gives,

$V_{ev}=\int d\textbf{r}d\textbf{R}\Phi^{\ast}(\textbf{R})\Phi(\textbf{R})\hat{V}_{ev}(\textbf{r}-\textbf{R})\Psi^{\dag}(\textbf{r})\Psi(\textbf{r})$

By means of the normal quantization steps, the e-v interaction follows,

$V_{ev}(\textbf{q})=\sum_{i,\sigma,\textbf{k},\textbf{q}}e^{i\textbf{q}\cdot \textbf{R}_{i}}\hat{V}_{ev}(\textbf{q})C^{\dag}_{\textbf{k}+\textbf{q},\sigma}C_{\textbf{k},\sigma}$

\noindent where $C^{\dag}_{\textbf{k},\sigma}$, $C_{\textbf{k},\sigma}$ are the creation and destruction operators for the electron with momentum $\textbf{k}$ and spin $\sigma$, respectively. The effective vacancy-vacancy (v-v) interaction $\hat{V}_{vv}(\textbf{q})$ in the momentum space could be built after eliminating the electron operators (see Fig.\ref{vv}(b)),

$\hat{V}_{vv}(\textbf{q})=\sum_{\textbf{k}}\frac{2\hat{V}^{2}_{ev}(\textbf{q})}{\varepsilon_{\textbf{k}+\textbf{q}}-\varepsilon_{\textbf{k}}}f(\varepsilon_{\textbf{k}+\textbf{q}})[1-f(\varepsilon_{\textbf{k}})]$

\noindent where $f(\varepsilon_{\textbf{k}})$ is the Fermi-Dirac (F-D) distribution, and $\varepsilon_{\textbf{k}}$ is the kinetic energy for electrons. In the temperature range $k_{B}T\ll \varepsilon_{F}$, the v-v interaction could be approximated as,
\begin{equation}\label{v-v}
\hat{V}_{vv}(\textbf{q})\approx-\frac{\hat{V}^{2}_{ev}(\textbf{q})}{4\pi^{2}}(\frac{\hbar^{2}}{2m_{e}})^{-3/2}\varepsilon^{1/2}_{F}
\end{equation}


\noindent where $\varepsilon_{F}$ is the Fermi energy and $k_{F}$ the Fermi vector. Seen from eq.(\ref{v-v}), the v-v interaction exhibits the mutual attractive potential at the temperatures $k_{B}T\ll \varepsilon_{F}$. The attractive v-v interaction is rational, because the aggregation of two vacancies would reduce the energy and be more stable as suggested by the calculated result\cite{vikram}.

\noindent\textbf{B. spallation-second order phase transition}

To simplify the theoretical treatment, take the assumption that the vacancies stay in the equilibrium state and the quantum statistical theory could be utilized to describe the spallation. And the dynamic behaviors in the non-equilibrium state will be discussed quantitatively later.

In the Bloch representation, the whole Hamiltonian for the vacancies reads,
\begin{equation}\label{Hamil}
H=\sum_{\textbf{k}}\frac{1}{2}(\varepsilon_{\textbf{k}}(\epsilon)+\varepsilon_{n}(\epsilon))b_{\textbf{k}}^{\dag}b_{\textbf{k}}
+\sum_{\textbf{q},\textbf{k},\textbf{k}'}\hat{V}_{vv}(\textbf{q})b_{\textbf{k}+\textbf{q}}^{\dag}b_{\textbf{k}'-\textbf{q}}^{\dag}b_{\textbf{k}'}b_{\textbf{k}}
\end{equation}
\noindent where $\varepsilon_{\textbf{k}}=(\hbar \textbf{k})^{2}/(2M_{v})$ is the kinetic energy for vacancies, where $M_{v}$ is the effective mass for the vacancy, about $10^2 $ times of the electron mass $m_{e}$ because of $\varepsilon_{\Lambda}\approx 0.01$ $  eV$($\Lambda=\pi/a$ is the reciprocal wave vector), the typical vibration energy for atoms, i.e. phonon energy. $\varepsilon_{n}(\epsilon)$ is the strain $\epsilon$ dependence of excitation energy ($\sim$eV ) for the vacancy. And for the kinetic energy of vacancies, the relation $\varepsilon_{\textbf{k}}(\epsilon)=\varepsilon_{\textbf{k}+\Lambda}(\epsilon)$ holds due to the transfer symmetry in the crystal. The prepositive factor 1/2 in the first term originates from the denominator 2 in the distribution $e^{-(\varepsilon_{\textbf{k}}(\epsilon)+\varepsilon_{n}(\epsilon))/2k_{B}T}$. $b^{\dagger}_{\textbf{k}}$,$b_{\textbf{k}}$ signify the creation and destruction operators for the vacancy with wave vector $\textbf{k}$. They are the boson operators and obey the Bose-Einstein (B-E) distribution. Here it is reasonable to take the vacancy field as the boson field, because the B-E distribution and the Boltzmann distribution are close to each other for the high excitation energy $\varepsilon_{n}(\epsilon)$.

So in the coherent state representation, the corresponding partition function for the vacancies in the equilibrium state reads,

$Z=\int D\left(\phi^{\ast}(\textbf{k})\phi(\textbf{k})\right)e^{-\int \phi^{\ast}(\textbf{k})\frac{1}{2}(\bar{\varepsilon}_{\textbf{k}}(\epsilon)+\bar{\varepsilon}_{n}(\epsilon))\phi(\textbf{k})\frac{d\textbf{k}}{(2\pi)^{3}}
-\int \phi^{\ast}(\textbf{k}+\textbf{q})\phi^{\ast}(\textbf{k}'-\textbf{q}) \bar{V}_{vv}(\textbf{q})\phi(\textbf{k}')\phi(\textbf{k})\frac{d\textbf{k}d\textbf{k}'d\textbf{q}}{(2\pi)^{9}}}$\\
\noindent where $\beta=1/(k_{B}T)$, $\bar{\varepsilon}_{\textbf{k}}(\epsilon)=\beta\varepsilon_{\textbf{k}}(\epsilon)$, $\bar{\varepsilon}_{n}(\epsilon)=\beta \varepsilon_{n}(\epsilon)$, $\bar{V}_{vv}(\textbf{q})=\beta \hat{V}_{vv}(\textbf{q})$.


 The effective interaction function $\bar{V}_{vv}(\textbf{q})$ may be expanded as
$\bar{V}_{vv}(\textbf{q})=\bar{V}_{vv}(0)+\bar{V}'_{vv}(0)\textbf{q} +\frac{1}{2}\bar{V}''_{vv}(0)\textbf{q}^{2}+\cdots$.
For simpleness, the first term in the expansion, independent of wave vector, is taken into account only. Using the random phase approximation, the self-energy $\Sigma(\textbf{k})$ for the vacancy gives,

$\Sigma(\textbf{k})\approx -\frac{\bar{V}_{vv}(0)}{3\pi^{2}}\frac{\Lambda^{3}}{\bar{\varepsilon}_{n}(\epsilon)}$



The achievement of the zero re-normalized excitation energy indicates the infinite correlation length among vacancies and the happening of phase transition and therefore the simultaneous critical behaviors caused by the coalescence of vacancies,

$\frac{1}{2}\bar{\varepsilon}_{n}(\epsilon)-\Sigma(\textbf{k})=0$

The transition temperature could be obtained,
\begin{equation}\label{T}
k_{B}T_{t}=-\frac{3\pi^{2}}{2\Lambda^{3}\hat{V}_{vv}(0)}\varepsilon_{n}^{2}(\epsilon)
\end{equation}
Substitute eq.(\ref{v-v}) into eq.(\ref{T}), it follows,

$k_{B}T_{t}=\frac{6\pi^{4}\varepsilon_{n}^{2}(\epsilon)}{\Lambda^{3}V_{ev}^{2}(0)}(\frac{\hbar^{2}}{2m_{e}})^{-3/2}\varepsilon_{F}^{-1/2}$

If the e-v interaction $V_{ev}(q)$ is taken as the screened coulomb interaction $\frac{4\pi Ze^{2}}{q^{2}+r_{es}^{-2}}$, where $r_{es}^{-2}$ is the electronic screening length for the ductile material, $Z$ is the effective charge for the vacancy, the transition temperature could be rewritten as

$k_{B}T_{t}=\frac{3}{8}\frac{4}{Z^{2}}(\frac{k_{F}}{\Lambda})^{3}\frac{\varepsilon^{2}_{n}(\epsilon)}{\varepsilon_{F}}$

This transition temperature means that the shocked temperature higher than $T_{t}$ would induce the complete spallation for the ductile metals.

This is the phase transition theory resorted to describe the spallation processes. It is like the ferromagnetic transition associated with the nucleation, the growth of clusters and the percolation of clusters, which are usually encountered in condensed matter physics. The  critical behaviors of the spallation have been discovered by the experimental results \cite{critical1, critical2}, which support the validity of the theory in this work.


The above theory is based on the equilibrium state for the spallation, not referring to the dynamic processes in the spallation, eg, the growth of voids. The following sections will focus on the dynamic properties in the spallation processes.

For the spallation process, when the catching rarefaction wave and the reflected rarefaction wave interconnect, there exist the zones centered at spallation planes with the large stress and strain gradient. These would lead to the dynamic properties and the time-evolution dependence of spallation. To describe them, the semi-classical Boltzmann Equation is employed,
\begin{equation}\label{Boltzmann}
\frac{\partial g_{\textbf{k}}}{\partial t}|_{coll}=\frac{\partial g_{\textbf{k}} }{\partial \textbf{k}}\frac{-v_{0}(\epsilon)}{\hbar}\nabla_{\textbf{r}}P(\textbf{r})+\frac{g_{\textbf{k}}\dot{\textbf{r}}}{2k_{B}T}\left[(\varepsilon_{\textbf{k}}(\epsilon)+\varepsilon_{n}(\epsilon))\frac{\nabla_{\textbf{r}} T(\textbf{r}(t))}{T}-\frac{\partial \varepsilon_{n}(\epsilon)}{\partial \epsilon}\nabla_{\textbf{r}}\epsilon(\textbf{r}(t))\right]
\end{equation}
\noindent where $g_{\textbf{k}}$ is the abbreviation of the temperature $T(\textbf{r})$, strain $\epsilon(\textbf{r})$ and momentum $\textbf{k}$ dependence of distribution function $g(\textbf{k},T(\textbf{r}),\epsilon(\textbf{r}))=e^{-(\varepsilon_{\textbf{k}}(\epsilon)+\varepsilon_{n}(\epsilon))/2k_{B}T}$ for the vacancies. In eq.(\ref{Boltzmann}), the first term on the right comes from the stress field accelerating the vacancies $\hbar \dot{\textbf{k}}\approx-v_{0}(\epsilon)\nabla_{\textbf{r}}P(\textbf{r})$, where $v_{0}(\epsilon)$ is the volume for one vacancy under strain and $P(\textbf{r})$ is the stress field.

Using the relaxation time approximation, $\partial g/\partial t|_{coll}=-(g_{\textbf{k}}-g^0_{\textbf{k}})/\tau_{\textbf{k}}=-g^{1}_{\textbf{k}}/\tau_{\textbf{k}}$, where $g^0_{\textbf{k}}$ is the equilibrium distribution for the vacancies and $\tau_{\textbf{k}}$ is the relaxation time for the vacancies with momentum $\textbf{k}$. Ignoring the phase coherence among the scattering, the relaxation time $\tau_{\textbf{k}}$ could be approximated as $1/\tau_{\textbf{k}}=1/\tau_{b\textbf{k}}+1/\tau_{v\textbf{k}}$, where $1/\tau_{b\textbf{k}}$ is the scattering rate induced by the scattering between the vacancies and the crystalline boundaries, and $1/\tau_{v\textbf{k}}$ is attributed to the scattering among the vacancies. The scattering rate for the vacancies with the velocity $\textbf{v}_{\textbf{k}}$ in the poly-crystalline metals with the typical grain sizes $L\sim 100$ $um$ is $1/\tau_{b\textbf{k}}\approx \textbf{v}_{\textbf{k}}/L$. And the related vital parameter, the elastic free length, for the vacancies is $l_{\textbf{k}}=\textbf{v}_{\textbf{k}}\tau_{v\textbf{k}}$.

According to the Fermi golden rule, the relaxation time $\tau_{v\textbf{k}}$ determined by the Umklapp scattering processes follows,

$\frac{-g^{1}_{\textbf{k}}}{\tau_{v\textbf{k}}}=\frac{2\pi}{\hbar}\sum_{\textbf{k}',\textbf{q}}|\hat{V}_{vv}(\textbf{q})|^{2}[g_{\textbf{k}}g_{\textbf{k}'}(1+g_{|\textbf{k}+\textbf{q}|-\Lambda})(1+g_{\textbf{k}'-\textbf{q}})
-g_{|\textbf{k}+\textbf{q}|-\Lambda}g_{\textbf{k}'-\textbf{q}}(1+g_{\textbf{k}})(1+g_{\textbf{k}'})
]\Theta(|\textbf{k}+\textbf{q}|-\Lambda)\delta(\varepsilon_{\textbf{k}}+\varepsilon_{\textbf{k}'}-\varepsilon_{|\textbf{k}+\textbf{q}|-\Lambda}-\varepsilon_{\textbf{k}'-\textbf{q}})$

\noindent where the function $\Theta(x) $ satisfies $\Theta(x)=\{^{1, for x\geq 0}_{0, others}\}$. Here the contributions to the vacancy scattering from other processes are not considered, eg. the scattering between dislocations and vacancies, the new generation and simultaneous re-combination of both the vacancies and interstitial atoms and so forth.
The vacancy current density(VCD) $j$ may be given,
\begin{equation}
j=\int \frac{\textbf{v}_{\textbf{k}}\tau_{\textbf{k}}g^0_{\textbf{k}}}{2k_{B}T}\left[(\varepsilon_{\textbf{k}}(\epsilon)+\varepsilon_{n}(\epsilon))\frac{\textbf{v}_{\textbf{k}}\cdot\nabla_{\textbf{r}} T(\textbf{r}(t))}{T}-\frac{\partial \varepsilon_{n}(\epsilon)}{\partial \epsilon}\textbf{v}_{\textbf{k}}\cdot\nabla_{\textbf{r}}\epsilon(\textbf{r}(t))+v_{0}(\epsilon)\textbf{v}_{\textbf{k}}\cdot\nabla_{\textbf{r}}P(\textbf{r})\right]\frac{d\textbf{k}}{(2\pi)^{3}}
\end{equation}
\noindent and $\textbf{v}_{\textbf{k}}\cdot\nabla_{\textbf{r}}P(\textbf{r})$ approximates $B_{0}\dot{\epsilon}$, where $B_{0}$ is the Young's modulus and $\dot{\epsilon}$ is the tensile strain rate.

For directional VCD, the pressure gradient may be governing. And at the temperature zone $\sim 10^3$ $K$, much lower than the characteristic excitation energy $\varepsilon_{n}(\epsilon)$, the scattering among vacancies may be quite weak and consequently the elastic free length $l_{\textbf{k}}$ may be longer than the grain sizes $L$, thereby making the relaxation time $\tau_{\textbf{k}}$ be determined by the crystalline sizes. So
\begin{equation}\label{current}
j\approx\frac{Lv_{0}(\epsilon)B_{0}\dot{\epsilon}}{6k_{B}T}\frac{\Lambda^{3}}{6\pi^{2}}e^{-\frac{\varepsilon_{n}(\epsilon)}{2k_{B}T}}
\end{equation}
 Seen from eq.(\ref{current}), the VCD is notably sensitive to the excitation energy $\varepsilon_{n}(\epsilon)$ and the temperature, and is proportional to the tensile strain rate. It also hinges on the sizes of grains, i.e. the smaller the sizes, the lower the VCD and vice versa. If take the typical parameters $T\sim 10^{3}$ $K$, $\varepsilon_{n}(\epsilon)\sim 1$ $eV$ and $L\sim 10^{-4}$ $m$, the VCD could arrive at $10^{28}$ $/(m^{2}\cdot s)$. The VCD, affected by these factors, will determine the damage evolution rate and the related properties of voids discussed in the following.

\noindent\textbf{C. damage evolution rate }

For the spallation planes, the damage evolution rate $\dot{V}$ could be given by the VCD,
\begin{equation}\label{damage rate}
\dot{V}=2jSv_{0}(\epsilon)
\end{equation}
\noindent where $S$ is the cross section of spallation plane and the factor 2 in the right term originates from the two series of rarefaction waves. Substitute eq.(\ref{current}) into eq.(\ref{damage rate}), it is
\begin{equation}\label{damagerate}
\dot{V}=\frac{LSv^{2}_{0}(\epsilon)B_{0}\dot{\epsilon}}{3k_{B}T}\frac{\Lambda^{3}}{6\pi^{2}}e^{-\frac{\varepsilon_{n}(\epsilon)}{2k_{B}T}}
\end{equation}
According to eq.(\ref{damagerate}), the damage evolution rate rely on the tensile strain rate and the sizes of the grains. The larger grain size and tensile strain rate would lead to a higher damage evolution rate, which have been totally confirmed by the experimental results\cite{Escobedo,xiaoyang}.

\noindent\textbf{D. stress on the boundary of void}

The large number of vacancies in the spallation planes would coalesce. The coalescence of vacancies will release considerable energy equivalent to the work done by the expanding boundary of void. So, for the stably growing void, the stress on the boundary of void $P_{b}$ is
\begin{equation}\label{boundarystress1}
P_{b}=\frac{1}{2}\frac{\varepsilon_{n}(\epsilon)}{v_{0}(\epsilon)}
\end{equation}
\noindent According to eq.(\ref{boundarystress1}), the stress $P_{b}$, independent of size and growing velocity of voids, is one intrinsic characteristic parameter for the specific ductile metal and is about $\sim 10$ $GPa$. For the metals with the fracture toughness $K_{IC}$ around the void, the critical radius $R_{c}$ for the stably growing voids is

\begin{equation}\label{criticalsize}
R_{c}=\frac{1}{\pi}(\frac{K_{IC}}{P_{b}})^{2}
\end{equation}
The existence of critical radius $R_{c}$ means that once the radius of void approach $R_{c}$ the cracks may occur and may induce the percolation among the nearby voids, which may be supported by the size distribution of voids in the experiments\cite{xiaoyang2}.

\noindent\textbf{E. plastic zone around the void }

\begin{figure}[B]
\includegraphics[scale=1,angle=0]{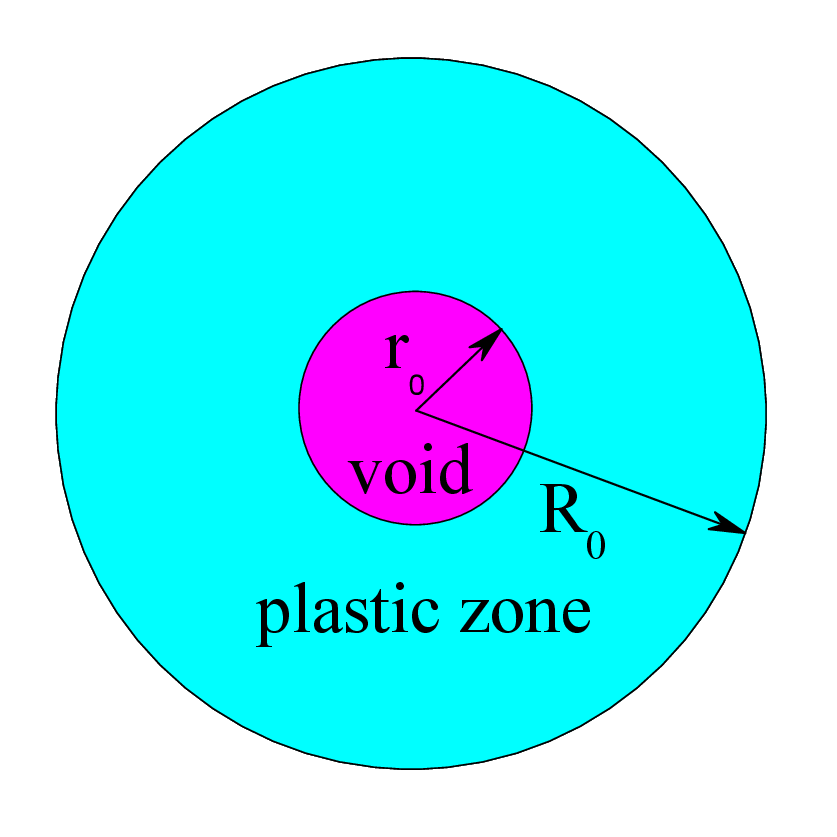}
\caption{Schematic diagram of the void (pink circled) with the radius $r_{0}$ and the surrounding plastic zone (light blued) with the radius $R_{0}$.}\label{plasticzone}
\end{figure}

For the monatomic ductile metals, the value of stress $P_{b}$ generally exceeds the yield strength of ductile metals, thereby generating a plastic shell surrounding the voids as shown in Fig.\ref{plasticzone}. The outer radius $R_{0}$ of the plastic shell could be obtained in terms of the simple calculation \cite{Reina} which is reasonable here due to the relatively small radial velocity,

$P_{r}(r)=A+\frac{B}{r^{3}}$ and $P_{\theta}(r)=A-\frac{B}{2r^{3}}$

\noindent where $P_{r}(r)$, $P_{\theta}(r)$ are the radial stress and tangential stress in the plastic zone, respectively. $A$ and $B$ are the parameters to be determined. By using the boundary conditions,$P_{r}(r_{0})=P_{b}$, $P_{r}(R_{0})-P_{\theta}(R_{0})=Y$, where $Y$, $r_{0}$ are the yield strength and the radius of void. The outer radius of the plastic shell could be obtained by noting the fact $P_{b}\gg Y$ for ductile metals,
\begin{equation}\label{plastic radius}
R_{0}=\sqrt[3]{\frac{3P_{b}}{2Y}} r_{0}
\end{equation}
 Here the calculation for the void in the equilibrium state is reasonable, because the expanding velocity of void is much smaller than the sound velocity and thus the stress could be regarded to be in the equilibrium state.

Substitute eq.(\ref{boundarystress1}) into eq.(\ref{plastic radius}), it is

$R_{0}=\sqrt[3]{\frac{3\varepsilon_{n}(\epsilon)}{4Yv_{0}}} r_{0}$

For the high purity aluminium, take the parameters\cite{aluminiumwiki,vacancy energy} $Y\sim 10$ $MPa$, $v_{0}= 16.7\times10^{-30}$ $m^{3}$,  $\varepsilon_{n}(\epsilon)\approx 0.65$ $eV$, the radius is calculated to be $R_{0}\approx 7.8 r_{0}$. Considering the work hardening effects, causing larger yield strength $Y$, the calculated result is in accord with the experimental results $\sim 4 r_{0}$ from the nano-indentation hardness experiments within the plastic zone around the void \cite{J. Belak,Belak}.


\noindent\textbf{F. nucleation and growth of void}

By clarifying the stress and plastic zone surrounding the void, the attention is paid to the nucleation rate and the growth velocity of void. The nucleation and the growth of voids with different sizes are the sources of damage evolution for the spallation process,
\begin{equation}\label{growthrate}
\dot{V}=\sum_{i}[\frac{dn_{i}}{dt}\frac{4\pi}{3}r^{3}_{i}+n_{i}4\pi r^{2}_{i}\frac{dr_{i}}{dt}]
\end{equation}
\noindent where $n_{i}$ is the number of voids with radius $r_{i}$. For the incipient spallation process, the nucleation may play the main role for the developing of damage. Substitute eq.(\ref{damagerate}) into eq.(\ref{growthrate}) and the nucleation for the nanometer void is given
\begin{equation}\label{nucleation}
\frac{dn_{i}}{dt}=\frac{LSv^{2}_{0}(\epsilon)B_{0}\dot{\epsilon} }{12k_{B}Tr^{3}_{i}}\frac{\Lambda^{3}}{2\pi}e^{-\frac{\varepsilon_{n}(\epsilon)}{2k_{B}T}}
\end{equation}
 In the following processes, the growth of voids may dominate the damage evolution process and thus the newly nucleation is subsidiary. The growth velocity for one void with radius $r_{0}$ could be obtained by considering that the vacancies propagating into the plastic zone will enter into the void in terms of plastic fluid or radial acceleration,
\begin{equation}\label{growthrate1}
\dot{r_{0}}\approx \frac{1}{2}jv_{0}(\epsilon)\sqrt[3]{(\frac{3P_{b}}{2Y})^{2}}
\end{equation}
Substitute eq.(\ref{current}) and eq.(\ref{boundarystress1}) into eq.(\ref{growthrate1}), and it is
\begin{equation}\label{growthrate2}
\dot{r_{0}}\approx \frac{Lv_{0}(\epsilon)B_{0}\dot{\epsilon}}{12k_{B}T}\frac{\Lambda^{3}}{6\pi^{2}}\sqrt[3]{\frac{9\varepsilon^{2}_{n}(\epsilon)v_{0}(\epsilon)}{16Y^{2}}}e^{-\frac{\varepsilon_{n}(\epsilon)}{2k_{B}T}}
\end{equation}

\noindent suggesting the proportionality to the tensile strain rate and the fixed growth velocity for the radius of voids regardless of their sizes. For the VCD and growth velocity of void, it should be noted that they strongly rely on the tensile strain rate and temperature. If the tensile strain rate arrive at $10^{6}$ $/s$, the VCD may achieve $\sim 10^{29}$ $/(m^{2}\cdot s)$ and may result in the growth velocity $\sim 100$ $m/s$. Actually, the growing void may link up the initial small voids in its plastic zone and form a larger void. If this is taken into account, the corrected expanding velocity $\dot{r}_{c}$ of void is

$\dot{r_{c}}(1-\int^{t}_{0}\frac{\dot{V}}{V_{0}}dt)\approx \frac{Lv_{0}(\epsilon)B_{0}\dot{\epsilon}}{12k_{B}T}\frac{\Lambda^{3}}{6\pi^{2}}\sqrt[3]{\frac{9\varepsilon^{2}_{n}(\epsilon)v_{0}(\epsilon)}{16Y^{2}}}e^{-\frac{\varepsilon_{n}(\epsilon)}{2k_{B}T}}
$

\noindent where $V_{0}$ is the initial undamaged volume in the spallation planes. In actuality, the damage evolution exhibits the time dependence, but for simpleness it is assumed to be constant and the corrected expanding velocity of void $\dot{r}_{c}$ follows,
\begin{equation}\label{growthratemodi}
\dot{r}_{c}= \frac{Lv_{0}(\epsilon)B_{0}\dot{\epsilon}}{12(1-\frac{\delta t}{t_{f}})k_{B}T}\frac{\Lambda^{3}}{6\pi^{2}}\sqrt[3]{\frac{9\varepsilon^{2}_{n}(\epsilon)v_{0}(\epsilon)}{16Y^{2}}}e^{-\frac{\varepsilon_{n}(\epsilon)}{2k_{B}T}}
\end{equation}
\noindent where $\delta$ is the final damage fraction in the spallation planes, equivalent to $\dot{V}t_{f}/V_{0}$ with $t_{f}$ as the duration time of whole spallation process. For the total spallation, the damage fracture is $\delta=1$.

 To estimate the growth rate of voids for tantalum and compare with the simulation result $\sim 200$ $m/s$ for the largest void\cite{critical behavior}, take the parameters\cite{vacancy energy} $\varepsilon_{n}(\epsilon)\sim 2.9 $ $eV$, $L\sim 16\times 53.3$ $\AA$, $T\sim 3\times 10^3$ $K$,
the tensile strain rate $\dot{\epsilon}\sim 2.7\times 10^{10}$ $/s$ estimated from the simulation for the largest void\cite{critical behavior}, the volume of one vacancy $v_{0}(\epsilon)\approx 18\times 10^{-30}$ $ m^{3}$, the yield strength $Y= 0.77$ $GPa$. The estimated growth velocity $\dot{r}_{c}\sim 130$ $m/s$, which agrees with the simulation result $\sim 200$ $m/s$.

\noindent\textbf{G. estimation of temperature}

To obtain the temperature field distribution around the void, the equation of energy conservation gives,
$V(r)C_{v}dT(r,t)=dU(r)=dQ(r)-P(r)dV(r)$

\noindent where $V(r)$ is the volume at the radius $r$ surrounding the void, $C_{v}$ is the specific heat capacity, $dU(r)$ is the variation of internal energy, and $dQ(r)$ the transport thermal energy and $-P(r)dV(r)$ the work done.
Through simple derivation, the time and radius dependence of temperature field $T(r,t)$ around the void obey
\begin{equation}\label{temperaturefield}
2\frac{P_{b}r^{3}_{0}}{r^{4}}\frac{dr}{dt}=C_{v}\frac{\partial T(r,t)}{\partial t}-K\frac{\partial^{2} T(r,t)}{\partial r^{2}}-\frac{2}{r}K\frac{\partial T(r,t)}{\partial r}
\end{equation}
where the radial velocity is\cite{stress wave} $dr/dt=\dot{r}_{0}r^{2}_{0}/r^{2}$, and $K$ is the thermal conductivity. Here only the rough estimation of the temperature for the boundary of void is done. The temperature field is thought to be in the stable state due to the much larger thermal transport velocity than the expanding velocity of void $\dot{r}_{0}$ and $ \partial^{2} T(r,t)/\partial r^{2}$ is assumed to be small. So the radius dependence of temperature field is

$T(r)\approx T(R_{0})+\frac{P_{b}r^{5}_{0}\dot{r}_{0}}{4K}(\frac{1}{r^{4}}-\frac{1}{R^{4}_{0}})$

Substitute the thermal conductivity for aluminium\cite{handbook} $K\approx 240$ $W/(m\cdot K)$, $r_{0}\sim 40$ $um$, the strain rate $\sim 10^{5}$ $/s$ and thus $\dot{r}_{0}\sim 10 $ $m/s$ into this equation, the obtained temperature on the boundary is $T(r_{0})\approx T(R_{0})+1200$ $K$. And the temperature decreases quickly with the increase of the distance from the void. Although the estimation is rough, it shows the much higher temperature around the void than other areas, which could  melt the aluminium. This is confirmed by the experimental results for the shocked pure aluminium, the recrystallization in a small region close to the void surface\cite{Belak}.

In the above theoretical treatment and discussions, the anisotropy, especially the crystalline orientations for single crystal and grain boundaries for the polycrystal, which may affect the strain rates and the effective mass of vacancies, is not considered. And the grain boundaries may have the trend to situate the vacancies and voids, thereby easy to form the crack along the grain boundaries, causing the spallation afterwards.

\noindent\textbf{3. Conclusions}

In summarized, the spallation in ductile metals is discussed in terms of phase-transition theory and non-equilibrium transport theory. Through the theoretical analysis, the spallation processes may be monitored by the diffusion and agglomeration of vacancies, and several points are concluded. (1) There exist the critical behaviors in the spallation processes. (2) A larger grain size of polycrystal and tensile strain rate and higher temperature will cause a larger damage evolution rate and larger void growth velocity. (3) The characteristic size for the voids and the characteristic stress on the void boundary exist, which rely on the vacancy excitation energy and the average volume occupied by the metal atom. (4) The stress fields and temperature fields around the void are obtained, and they are high near the void and may cause the melting, but they decrease rapidly as the distance away from the void increases. (5) The plastic zone surrounding the formed voids is obtained and found to be governed by the characteristic stress. (6) The void growth in the spallation processes may arise from the coalescence of vacancies rather than the emitting of dislocations, and the emitted dislocations around the void is the result of growing void, not the origins.
Through comparisons,  the obtained theoretical results are in agreement with the experiments and the simulations.

\noindent\textbf{4. Acknowledgements}

The author is grateful to colleagues Guangfu Ji, Hongliang He, Chuanmin Meng and Liang Wang for the helpful discussions and the amounts of provided information.

\end{document}